\documentclass[aps,showpacs,twocolumn,prb]{revtex4}
\usepackage{amsfonts}
\usepackage{amsmath}
\usepackage{amssymb}
\usepackage{graphicx}

\newcommand{\placefigure}[4]
{
 \begin{figure}[t]
 \includegraphics[width=#2]{#1}
 \caption{#3}
 \label{#4}
 \end{figure}
}

\begin{document}
\title{Ordering of geometrically frustrated classical and quantum
  Ising magnets} \author{Ying Jiang}
\affiliation{Institut f\"ur Theoretische Physik, Universit\"at zu
  K\"oln, Z\"ulpicher Strasse  77, 50937 K\"oln, Germany}
\author{Thorsten Emig}
\affiliation{Institut f\"ur Theoretische Physik, Universit\"at zu
  K\"oln, Z\"ulpicher Strasse 77, 50937 K\"oln, Germany}

\begin{abstract}
  A systematic study of both classical and quantum geometric
  frustrated Ising models with a competing ordering
  mechanism is reported in this paper. The ordering comes in the
  classical case from a coupling of  2D layers and in the quantum
  model from the quantum dynamics induced by a transverse field.  By
  mapping the Ising models on a triangular lattice to elastic lattices
  of non-crossing strings, we derive an exact relation
  between the spin variables and the displacement field of the
  strings.  Using this map both for the classical (2+1)D stacked model
  and the quantum frustrated 2D system, we obtain a microscopic
  derivation of an effective Hamiltonian which was proposed before on
  phenomenological grounds within a Landau-Ginzburg-Wilson approach.
  In contrast to the latter approach, our derivation provides the
  coupling constants and hence the entire transverse
  field--versus--temperature phase diagram can be deduced, including
  the universality classes of both the quantum and the
  finite--temperature transitions. The structure of the ordered phase
  is obtained from a detailed entropy argument. We compare our
  predictions to recent simulations of the quantum system and find good
  agreement. We also analyze the connections to a dimer model on the
  hexagonal lattice and its height profile representation, providing a
  simple derivation of the continuum free energy and a physical
  explanation for the universality of the stiffness of the height
  profile for anisotropic couplings.
\end{abstract}

\maketitle

\section{Introduction}

Frustrated two-dimensional magnetic systems exhibit a rich variety
of phases and critical points. The consequences of quantum or
thermal fluctuations about the highly degenerate ground states of
frustrated magnets are important to understand in the ongoing
quest to find novel exotic quantum states \cite{Sachdev-book}.
The past decades have seen a resurgence of significant interest in
a systematic study of geometrically frustrated antiferromagnets
\cite{Diep-book}.  Geometric frustration arises in materials
containing antiferromagnetically coupled moments which reside on
geometrical units, that inhibit the formation of a collinear
magnetically ordered N{\'e}el state, and induces a macroscopically
large degeneracy of the classical ground state.

For magnets with a discrete Ising symmetry, the complexity of the
ground-state manifold can endow the system with a continuous
symmetry. Such symmetry is of particular importance to 2D quantum
magnets at low temperatures since then the Mermin-Wagner theorem
applies, precluding an ordered phase \cite{Mermin+66}. Another
possible but contrary scenario is ``order-from-disorder''
\cite{Villain+80} where zero-point fluctuations select a small
class of states from the ground-state manifold since those states
are particularly susceptible to fluctuations. This fundamental
mechanism can produce an ordered symmetry-broken state
\cite{moessner1}.  Hence one expects weak competing fluctuations
about the classical ground states to be able to generate new
strongly correlated states and (quantum) phase transitions of
unexpected universality classes.

From a theoretical perspective, it is natural to try to understand
the role of frustration from extremely simple interactions and
dynamics. The possibly simplest realization of classical
frustration is found in the antiferromagnetic Ising model on a
triangular lattice (TIAF). Each elementary triangle is frustrated,
and the TIAF is disordered even at zero temperature with a finite
entropy density and algebraic decaying spin correlations
\cite{Wannier+Houtappel}. The effect of quantum fluctuations about
the highly degenerate ground states can be studied in its simplest
form by introducing quantum spin dynamics from a magnetic field
which is transverse to the spin coupling.  For this transverse
field TIAF model, and its companions on other 2D lattices,
Moessner and Sondhi have argued the existence of both ordered and
spin liquid phases \cite{moessner1}.  In the limit of a small
field, the quantum ground state is constructed as a linear
superposition of classical ground states which maximize the number
of spins which can be flipped to gain transversal field energy at
no cost in exchange energy. This yields a strong suppression of
configurations and, since the TIAF is already critical at zero
field, order emerges.  As a result, there will be a discontinuity
in entropy and correlations in the ground state at vanishing field
and zero temperature. At a large transverse field, the TIAF has a
gaped paramagnetic ground state.  It is separated from the ordered
state by a quantum critical point. The additional effect of
thermal fluctuations has been studied quantitatively so far only
in simulations \cite{Isakov+03}.

Experimental realizations of these frustrated Ising systems can be
found either directly in magnets with strong anisotropy, e.g., in
LiHoF$_4$ \cite{Aeppli+98}, or indirectly (via the equivalent
(2+1)D classical model) in stacked triangular lattice
antiferromagnets \cite{Collins+97} with strong couplings along the
stacking direction as studied in recent experiments on CsCoBr$_3$
\cite{Mao+02}.  However, transverse field Ising models can also
provide insight into some more complicated systems in certain
limits.  They may describe the singlet sector below the spin gap
of frustrated antiferromagnetic quantum Heisenberg models, e.g.,
on the Kagome lattice, since the latter model can then be
formulated as a ${\mathbb Z}_2$ gauge theory which in turn is
related by duality to the transverse field Ising system on the
dual lattice of the original Heisenberg model \cite{Nikolic+03}.
An ordered Ising phase describes then in the Heisenberg problem a
paramagnet with spin waves forming the gaped excitations. A
related approach to study Heisenberg antiferromagnets is via
quantum dimer models (QDM) on the same lattice \cite{Rokhsar+98}.
Depending on the lattice symmetry, the QDM exhibits a disordered
state (on the triangular lattice)
\cite{Moessner+01a,Moessner+01b}, corresponding to a resonating
valence bond or spin-liquid phase, or it is found always in
ordered valence-bond solid phases (on the hexagonal lattice)
\cite{Moessner+01a,Moessner+01b}.  Interestingly, the spin liquid
phase renders the triangular QDM a promising candidate for quantum
computing \cite{Ioffe+02}.  For Ising spins, there is also a
direct correspondence between QDM's with just kinetic terms and
fully frustrated Ising models in a small transverse field on the
dual lattice \cite{moessner1}.  In particular, the transverse
field TIAF maps to the hexagonal QDM.

For comparison to the new developments presented in this work, we
briefly review the salient achievements for the classical stacked and
quantum TIAF obtained in earlier studies.  Symmetry arguments have
been used to guess a Landau-Ginzburg-Wilson (LGW) theory for the
(2+1)D classical stacked system \cite{Blankschtein+84}.  After
translation to the quantum TIAF, this approach suggests a quantum
critical point of 3D XY universality at intermediate transverse field
strength and an extended critical phase at finite temperatures
\cite{moessner1}.  However, the Suzuki-Trotter mapping relating the 2D
quantum and (2+1)D classical systems involves a scaling limit with a
diverging anisotropy of the classical exchange coupling which may
invalidate a LGW approach \cite{suzuki1}. Moreover, the latter
approach has been put somewhat into question, mainly since it has been
argued that it fails to describe the (classical) system at low
temperatures since it neglects the restriction to classical spin
values $\pm 1$ and thus geometrical frustration \cite{Coppersmith85}.
Very recent Monte Carlo simulation support the LGW based conjecture
for the phase diagram but the actual computations were performed for
the (2+1)D classical problem \cite{Isakov+03}. Recently, the weak
field behavior has been also studied in terms of a quantum kink
crystal \cite{Mostovoy+03}.

In a recent letter \cite{Jiang+05} the present authors have
presented a string description of the transverse field TIAF in
order to obtain a quantitative prediction for the phase diagram.
In the present paper, we will give a full account of the relation
between spins and non-crossing strings, including a more complete
presentation of the relation to dimer models and its surface
height representation.  We exploit and extend the quantum dimer
analogy in order to map the TIAF, at arbitrary transverse field
strength, to quantum strings which result from the superposition
of dimer configurations plus defects and a fixed reference dimer
state. Using the Suzuki-Trotter theorem \cite{suzuki1} we obtain a
stack of coupled 2D layers of classical strings which in the
Suzuki-Trotter limit of infinitely many layers can be described by
the Villain model \cite{Villain75}.  We show that the LGW action
\cite{Blankschtein+84}, which was employed in the vicinity of the
quantum critical point \cite{moessner1}, can be derived
microscopically from the quantum string action if the phase of the
complex LGW order parameter is identified with the displacement
field of the string lattice.  However, our approach explicitly
takes into account frustration which is encoded in the topological
constraint on the phase field resulting from the non-crossing
property of the strings.  This constraint restricts the phase
field configurations of the order parameter and thus distinguishes
our theory from the original LGW approach.  Our approach confirms
explicitly the 3D XY universality of the quantum critical point,
and allows to predict the phase diagram at {\it arbitrary}
transverse field strength and finite temperature. Using an entropy
argument for the strings, we can determine the nature of the
ordered phase.

In this paper we present first a thorough discussion of the
relations between Ising spins on the triangular lattice, dimers on
the hexagonal lattice and its height representation and elastic
string lattices. After a introduction of the spin models in
Section \ref{sec:models}, we explore the relations to dimers and
strings in Section \ref{sec:classical_2D} for the classical 2D
Ising model on the triangular lattice in order to set up the
formalism for the following sections. In Section \ref{sec:stacked}
we map the Ising model on a stacked triangular lattice to (2+1)D
string lattice which is described by a 3D XY model with a 6-fold
symmetry breaking term. Using the results of the latter section,
we perform a Suzuki-Trotter mapping of the 2D quantum Ising system
to the classical stacked model which enables us to predict a
quantitative phase diagram for the quantum frustrated model in
Section \ref{sec:quantum}. Finally, we conclude with a summary and
a discussion of potential extensions of our work in Section
\ref{sec:discussion}.

\section{Models}
\label{sec:models}

We are interested in two models which are both based on the Ising
antiferromagnet on a triangular lattice (TIAF). First, we study a
classical three dimensional Ising system which consists of a stack of
TIAF's which are coupled ferromagnetically. The Hamiltonian reads
\begin{equation}
\label{stackedH}
H_{\rm 3D}=J_\parallel\sum_{\langle
ij\rangle,k}\sigma_{ik}\sigma_{jk}-J_\perp\sum_{i,k}
\sigma_{ik} \sigma_{ik+1},
\end{equation}
with $\sigma_{ik}=\pm 1$ and where $J_\parallel$, $J_\perp>0$ and
$\langle ij \rangle$ indicates summation over nearest-neighbor
pairs in each TIAF plane. This system is fully frustrated in each
layer, but has no competing interaction along the stacking
direction. This model has been studied originally within a
Landau-Ginzburg-Wilson (LGW) approach by Blankschtein et al.
\cite{Blankschtein+84}.  In fact, much of the physics of the
stacked triangular lattice is similar to that of the
two-dimensional triangular lattice.  However, the presence of the
third dimension has the tendency to stabilizes ordered phases.
Indeed, for all well-characterized materials which have triangular
magnetic lattices, the long-range order at low temperatures is
three-dimensional in nature.  The stacked system is of direct
experimental relevance since it describes the low-temperature
physics of the Ising-like compounds CsCoCl$_3$, CsCoBr$_3$ and
related materials where transition metal atoms form chains along
the stacking direction which are coupled through three equivalent
halogen atoms. For a review on these compounds see
Ref.~\onlinecite{Collins+97}. Almost the entire existing body of
theory on this stacked system is based on the LGW approach,
mean-field theory and numerical simulations with sometimes
conflicting results for the low-temperature ordered phase.  We
will briefly review previous developments for this model in
Section \ref{sec:stacked}.

As shall be demonstrated below, the previously introduced model is
also useful in describing the two-dimensional quantum antiferromagnet
which results from applying a transverse magnetic field to the TIAF.
The latter system has the Hamiltonian
\begin{equation}
\label{eq:model} H=J \sum_{\langle i,j \rangle} \,\sigma^z_i
\sigma^z_j + \Gamma \sum_i \sigma^x_i \, ,
\end{equation}
$\sigma^x$, $\sigma^z$ are Pauli operators, and $\Gamma$ is the
transverse field. This model is of particular interest since it
combines a standard realization of geometric frustration with
simple quantum dynamics. The transverse field induces tunnelling
between the exponentially large number of classical ground states
at zero temperature.  One can argue that the quantum fluctuations
select a smaller susceptible class of the ground states. This
would lead then to a reduction of ground state entropy, and an
``{\it order from disorder}'' \cite{Villain+80} phenomenon is
expected.

Both the classical stacked magnet and the 2D quantum TIAF are related
to the quantum dimer model (QDM) on the hexagonal lattice.

\section{String picture of classical frustrated 2D magnets}
\label{sec:classical_2D}

\subsection{From spins to dimers}

\placefigure{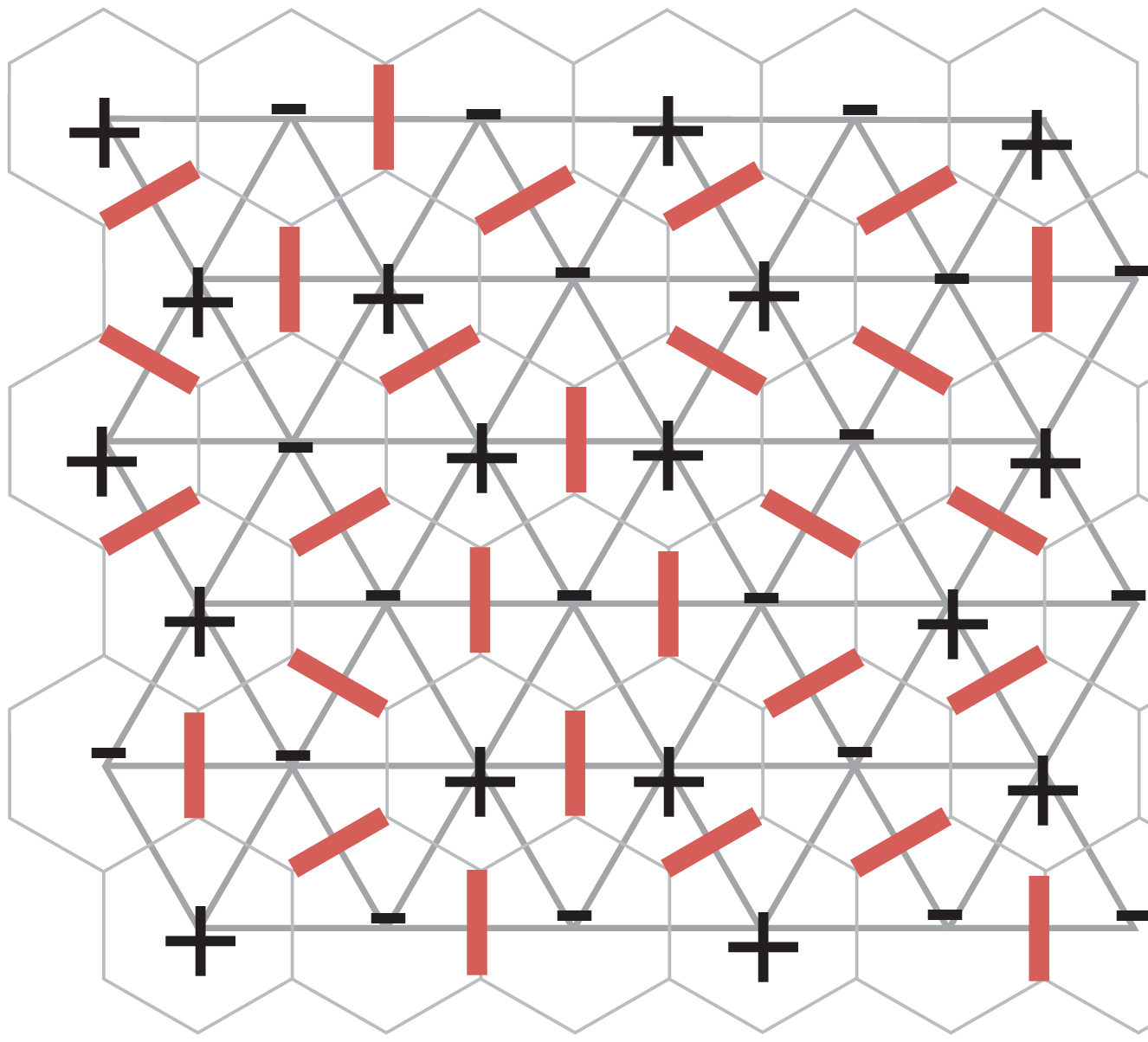}{0.7\linewidth}{(color online)
Mapping of a spin configuration to a dimer covering on the dual
hexagonal lattice. }{fig:dimer}

We start with the description of the antiferromagnetic Ising model on
the triangular lattice in order to explain its relation to dimers and
strings. The Hamiltonian is
\begin{equation}
\label{eq:tiaf-ham}
H_{\rm TIAF}=\sum_{\langle i,j\rangle} J_{ij}\sigma_{i} \sigma_{j}
\end{equation}
where the couplings $J_{ij}$ are equal to one of the three
positive constants $J_1$, $J_2$ or $J_3$ depending on the
orientation of the bond $(ij)$ relative to the three main lattice
directions. It is well known\cite{Wannier+Houtappel} that the
ground state of this system has exponentially large degeneracy in
the isotropic case $J_1=J_2=J_3$, leading to a finite entropy
density and the absence of order even at zero temperature. For all
ground states of the general model of Eq.~(\ref{eq:tiaf-ham}) each
triangle has exactly one frustrated bond, and hence there is a
one-to-one correspondence (modulo a global spin flip) of all spin
ground state configurations to complete dimer coverings of the
dual hexagonal lattice. This can be seen from
Fig.~\ref{fig:dimer}: If one places a dimer across every
frustrated bond, one obtains the corresponding dimer covering of
the dual hexagonal lattice, where each lattice site in the
hexagonal lattice is touched by one and only one dimer
\cite{Nienhuis+84}. If the two smallest couplings are equal,
$J_1=J_3 < J_2$, then dimers can occupy bonds in directions $1$
and $3$ only and there is still a huge ground state degeneracy.
But the entropy at $T=0$ scales now $\sim \sqrt{N}$ with the
number of lattice sites, $N$, due to the constraint that dimers
cannot touch each other. Thus the entropy density vanishes and the
system is ordered at $T=0$ but disordered at any finite
temperature.  (At finite $T$ topological defects have to be
considered, see below.) If one of the three coupling constants is
the smallest, then there is only one dimer ground state
configuration with all dimers being perpendicular to the direction
with smallest coupling, and order survives at finite temperature.

The analogy to dimers can be expressed by the fact that the Ising
partition function in the limit of vanishing temperature is
proportional to that of the dimers which reads
\begin{equation}
\label{eq:Z_dimer}
Z_D = \sum_{\{D\}} z_1^{n_1} z_2^{n_2}\, ,
\end{equation}
where the sum runs over all complete dimer coverings of the hexagonal
lattice, and $n_1$ and $n_2$ is the number of dimers being
perpendicular to direction 1 and 2, respectively.  Here we assumed
that $J_3 \le J_1$, $J_2$ so that the weights for the two types of
non-vertical bonds are given by
\begin{equation}
z_1=e^{-2(J_1-J_3)/T},\;z_2=e^{-2(J_2-J_3)/T},
\label{eq:bond-weight}
\end{equation}
and all vertical bonds have a weight of unity.  Of course, for the
TIAF the isotropic case with $J_1=J_2=J_3$ is of most interest but for
the dimer model itself anisotropic weights provide some interesting
insight as will become clear below. We note that even at finite
temperature the analogy to dimers will turn out to be useful but then
defects (triangles with all three bonds crossed by dimers) have to be
included.

\subsection{From dimers to strings}
\label{sec:dimers-to-strings}

\placefigure{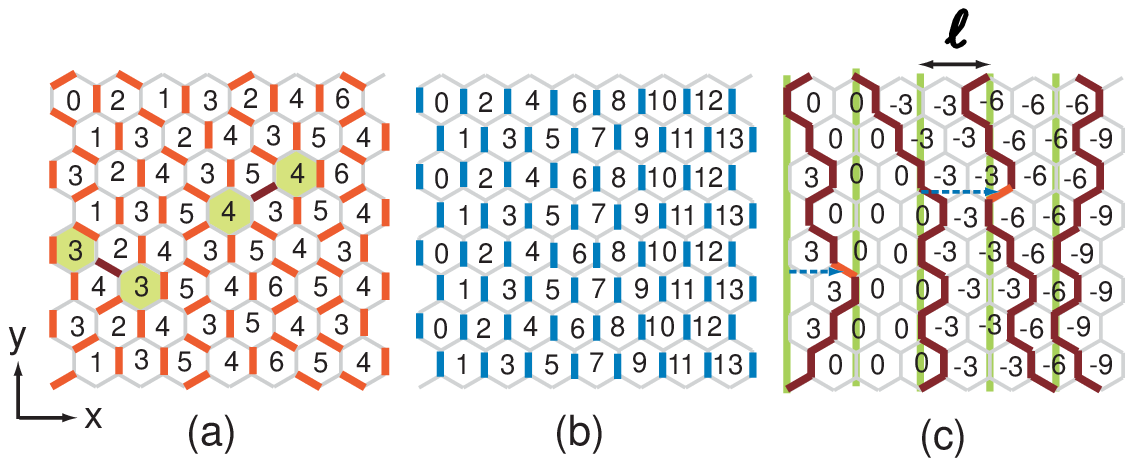}{1\linewidth}{(color online) Mapping a dimer
  covering (a) to a string configuration (c) via the reference state
  (b). The numbers denote the height profiles associated with the
  dimer coverings. The displacement of a string from its straight
  reference position is determined by the height profile at the two
  plaquettes which are joined by the displaced dimer in the original
  dimer covering (a).  }{fig:height-line}

In this subsection we make use of another mapping which relates
dimer configurations to fluctuating strings which are directed and
non-crossing \cite{yokoi+86}. This mapping applies to general
weights for the dimers but of primary interest is the TIAF at
$T=0$ where the dimer weights are either unity or zero, and there
is a one-to-one correspondence (modulo a global spin flip) of
Ising ground states to string configurations.  The string
representation results from the superposition of a given dimer
covering and a fixed reference dimer covering where all vertical
bonds of the hexagonal lattice are occupied by dimers, see
Fig.~\ref{fig:height-line}(b). The superposition is an "exclusive
or" operation, i.e., only if a given bond is covered by a dimer
either in the original covering or in the reference covering, it
will be covered in the superposition.  The resulting superposition
is no longer a dimer covering (since dimers touch each other) but
an array of strings which are directed (along the reference
direction) and non-crossing (due to the fact that each site is
touched by exactly one dimer in the original dimer covering)
\cite{zeng1}. The fluctuations of the lines result from the
non-zero entropy density of the spin system. Or in other words, we
map the zero temperature TIAF to a line lattice at a finite
(virtual) temperature $T_l$.

Before we parameterize the string configurations, it is useful to
introduce a height profile $h$ which can be associated with a
dimer covering. If the lattice covered by dimers is bipartite,
then there exists a well-defined (single-valued) height profile
which is defined on the sites of the original (triangular)
lattice.  Starting at an arbitrary site with some integer number,
one follows a triangle pointing down clockwise and changes the
height by +2 (-1) if a (no) dimer is crossed. Repeating the latter
process for all triangles pointing downward, one obtains a
consistent height on all sites, see Fig. \ref{fig:height-line}
(a). If one subtracts the height profile of the fixed reference
dimer covering from a given height profile, one observes that the
previously introduced strings separate domains of equal height
which is an integer multiple of $3$, see
Fig.~\ref{fig:height-line}(c). It will turn out that the height
profile measures the roughness, i.e., the displacement of the
strings from a perfectly straight configuration.

\placefigure{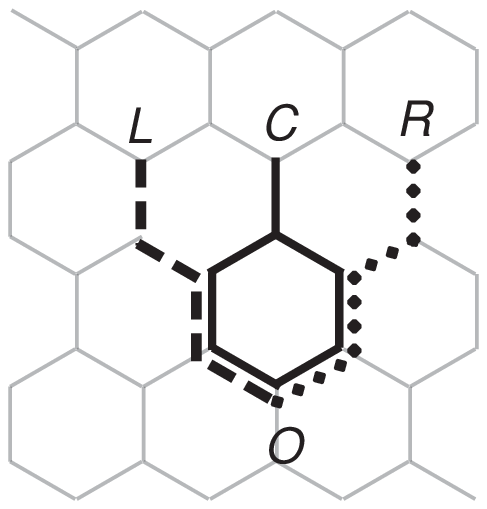}{0.4\linewidth}{Different kind of steps of the
  random walk of a string. }{fig:stiffness}

The array of strings is characterized only by its density
$\rho=1/\ell$ and the elastic line tension $g$ of the strings. The
interaction between the strings consists only of the non-crossing
constraint and thus introduces no additional energy scale. The
configurations of a single directed string are characterized by
its position function $x_{i}(z)$ since overhangs are forbidden.
The reduced energy of all strings is then
\begin{equation}
\frac{H_l}{T_l}= \sum_{i}\int dy\,  \frac g{2T_l}\left(
\frac{dx_{i}}{dy}\right) ^{2}.
\label{conline}
\end{equation}
First, we determine the line tension $g$ from a simple random walk
argument. If $\bar x$ denotes the mean string position, then the
total mean squared displacement of a string of length
$L=\sqrt{3}aM$ along the $y$-direction (after $M$ steps), with $a$
the triangular lattice constant, must be
\begin{equation}
\sum_{i=1}^M \langle (x_i - \bar x)^2 \rangle = LT_l /g
\end{equation}
in order to be consistent with the result of the continuum model of
Eq.(\ref{conline}). Here $x_i=-a,\;0,\;+a$ are the three possible
transversal steps.  The different kind of steps correspond to the
paths $OL$, $OC$ and $OR$ shown in Fig.~\ref{fig:stiffness}.  The
corresponding probabilities for the steps can be expressed in terms of
the weights of the occupied non-vertical bonds as
\begin{equation}
p_{L}=\frac{z_{2}^{2} }{(z_{1} +z_{2})^{2}}, \;p_{C}=\frac{2z_{1}
z_{2}}{(z_{1} +z_{2})^{2}}, \; p_{R}=\frac{z_{1}^{2}}{(z_{1}
+z_{2})^{2}}.
\end{equation}
The mean position after one step is
\begin{equation}
\bar x=(p_{R} -p_{L})\, a= \frac{z_{1} -z_{2}}{z_{1} +z_{2}}a.
\end{equation}
Thus, after $M$ steps, the variance of the transverse wandering
around the mean position is
\begin{eqnarray}
\sum_{i=1}^{M}\langle(x_{i} - \bar x)^{2}\rangle =L
\frac2{\sqrt{3}\, (2+\eta+\eta^{-1})}a,
\end{eqnarray}
where $\eta=z_{1}/z_{2}$, yielding the line tension
\begin{equation}
\frac g{T_l} = \frac{\sqrt{3} \, (2+\eta+\eta^{-1})} {2a}.
\label{stiffness}
\end{equation}
The {\it mean} density of strings depends also on the dimer weights of
the non-vertical bonds. It can be computed exactly with the
result \cite{Nienhuis+84,yokoi+86}
\begin{equation}
\rho=\frac2{\pi a} \arcsin\left[ \frac{(z_1+z_2)^2 - 1}{4z_1
  z_2}\right]^{1/2},
\label{density}
\end{equation}
for $z_1+z_2>1$.  For $z_1+z_2 \le 1$ the density vanishes
\cite{kasteleyn1}, corresponding to a Kasteleyn or
commensurate-incommensurate transition. For the isotropic TIAF,
one has $z_1=z_2=1$ and $\rho=2/(3a)$.

Before we introduce the relation between the string displacement
and the height profile, and a corresponding continuum description,
it is instructive to focus on the free energy of the string array.
This will allow for an independent check of our random walk
description. Moreover, from the free energy one can determine the
elastic constants on large length scales which are needed for the
proper continuum model.

Actually, the density of strings is determined by the densities
$\rho_1$, $\rho_2$ of non-vertical dimers with weights $z_1$, $z_2$,
respectively. This can be seen from the fact that along a cut
perpendicular to the string direction each non-vertical dimer
corresponds to one string.  It follows from Eq.~(\ref{eq:Z_dimer})
that
\begin{equation}
\rho_{i}=\frac{z_{i}}{N}\frac{\partial\log Z_D}{\partial z_{i}}\, ,
\end{equation}
for $i=1$, $2$, where the number of triangular lattice sites, $N$,
equals the total number of dimers. The string density then reads
\begin{equation}
\rho = \frac{\rho_1+\rho_2}{a}\, .
\end{equation}
Now we change variables from $z_1$, $z_2$ to $\eta=z_1/z_2$ and
$\tilde\rho = \rho a$, see Eq.(\ref{density}). Since $z_1 \partial
\eta/\partial z_1 + z_2 \partial \eta/\partial z_2 =0$, and using
Eq.(\ref{density}), we have
\begin{equation}
z_1 \frac{\partial}{\partial z_1} + z_2 \frac{\partial}{\partial z_2} =
\frac{1+\eta^2+2\eta \cos(\pi \tilde \rho)}{\pi \eta \sin(\pi\tilde\rho)}
\frac{\partial}{\partial \tilde \rho} \, .
\end{equation}
This allows us to derive a differential equation for the free energy
density of the dimers,
\begin{equation}
f_{D} = - \frac{2 \log Z_D}{\sqrt{3} N a^2}
\end{equation}
as a function of the parameters $\eta$, $\rho$ of the string
system. $f_D$ fulfills the equation
\begin{equation}
\frac{\partial f_D}{\partial \rho} = - \frac{2\pi}{\sqrt{3}}
\frac{\eta \rho \sin (\pi a \rho)} {1+ \eta^2 + 2\eta \cos(\pi a
\rho) }.
\end{equation}
Integrating the latter expression yields the exact result for the
lattice string model. In the continuum limit $a \to 0$, the result
simplifies to
\begin{equation}
\label{eq:f_dimers}
f_D = - \frac{\pi^2}{3} \frac{2a}{\sqrt{3}(2+\eta+\eta^{-1})} \,\rho^3 \, .
\end{equation}
There is an alternative and simple approach to obtain the free energy
in the continuum limit. The strings can be regarded as the world lines
of free fermions in one dimension \cite{Pokrovsky+79}. The
non-crossing condition is then implemented automatically by the Pauli
principle. In the limit of infinitely long strings, the ground state
energy density of the fermions equals the free energy density of the
strings if the fermion mass is mapped to the line tension $g$ and
$\hbar$ to the temperature $T_l$. From the expression for the ground
state energy of one-dimensional free fermions follows immediately the
reduced free energy density of the strings,
\begin{equation}
\label{eq:f_strings}
f_l = f_1 \rho + \frac{\pi^2}{6} \frac{T_l}{g} \rho^3 \, ,
\end{equation}
where $f_1$ is the reduced free energy (per length) of a single
string, and the last term is generated by the reduction of entropy
due to the non-crossing condition. A direct comparison of the free
energies of Eqs.(\ref{eq:f_dimers}) and (\ref{eq:f_strings}) is
not possible since the number of strings varies with the dimer
coverings although the number of dimers is fixed. Only the {\it
mean} density of strings is fixed. Hence, one has to consider the
grand potential density $j$ for the string system which follows
easily from the free energy density,
\begin{equation}
\label{eq:j_potential}
j= f_l - \mu \rho = - \frac{\pi^2}{3} \frac{T_l}{g} \rho^3 \, ,
\end{equation}
where $\mu=\partial f_l/\partial \rho$ is the chemical potential
(per string length). Comparing Eq.(\ref{eq:f_dimers}) and
Eq.(\ref{eq:j_potential}), we see that $f_D=j$ exactly for the
expression of the line tension which we obtained above from an
independent random walk argument, see Eq.(\ref{stiffness}). Thus
in the continuum limit the ground states of the TIAF and the dimer
model can be described as free fermions with their mass determined
by our simple random walk argument on the lattice.

\placefigure{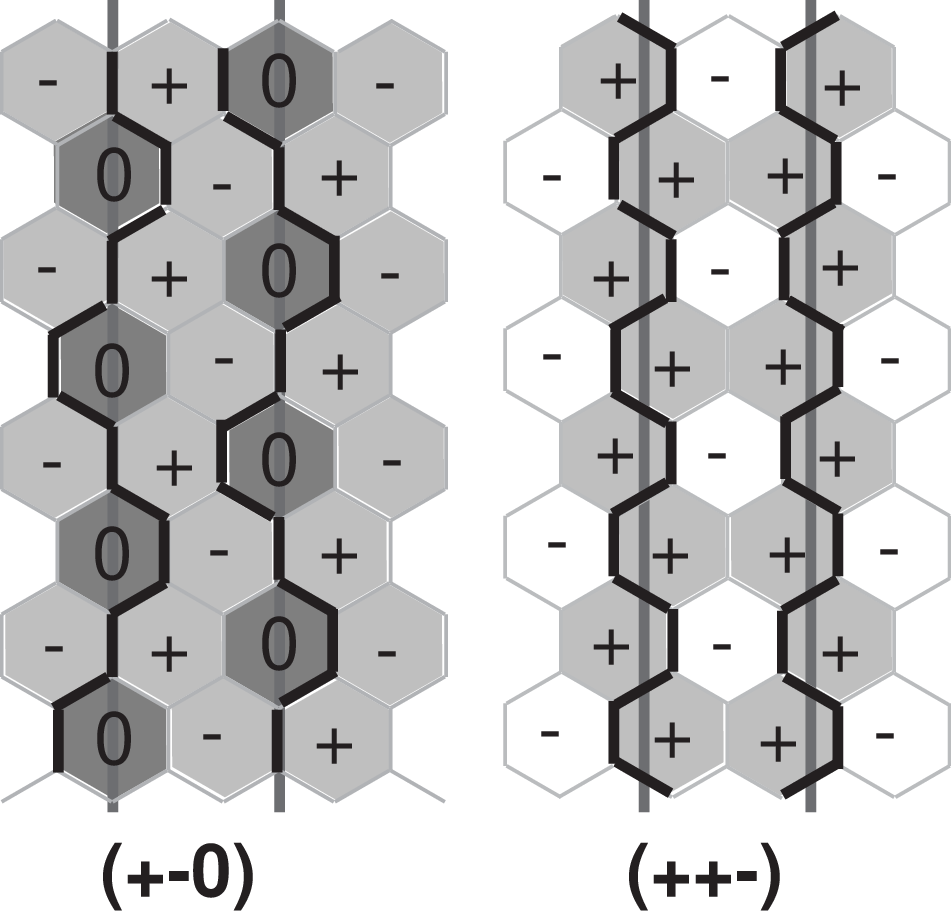}{0.6\linewidth}{Two different flat
  states, with the corresponding sublattice magnetizations of the
  isotropic TIAF indicated below.}{fig:flat}

Now we turn to the structural properties of the string lattice.  The
most ordered state consists of straight (``flat'') strings which do
not wander transversally. Depending on the string density there are
different commensurate and incommensurate states possible. For the
discussion of the flat states we focus on the state with a maximal
density of $\rho=2/(3a)$ which corresponds to the most interesting
case of isotropic couplings for the TIAF. In this case there are two
non-equivalent classes of flat strings which are not related by shifts
by a nearest-neighbor vector of the triangular lattice, see
Fig.~\ref{fig:flat}. These classes translate directly to different
ordered spin states which are characterized by their sublattice
magnetizations.  One state, $(+-0)$, is obtained by orienting two
sublattices uniformly but opposite while the third sublattice has the
same number of $+$ and $-$ spins, leading to zero magnetization.
This, on average, locks the straight strings {\it on} the sublattice
with zero magnetization. Since each of the three sublattices can be
chosen as the one with zero magnetization, equivalent flat string
states are related by a shift of $a/2$.  The other state, $(++-)$, has
up spins on two sublattices and down spins on the other sublattice
which places the non-vertical dimers on straight strings that are
locked symmetrically between the two sublattices with equal
magnetization.

The displacement $u$ from flat strings is determined by the
original height profile $h$ of the corresponding dimer covering.
Since only non-vertical dimers are conserved under the subtraction
of the reference covering, we will defined $u$ for each
non-vertical bond of the hexagonal lattice. If a bond is occupied
by a string segment, then $u$ measures the normal distance between
the corresponding flat string and the center of the segment,
cf.~Fig.~\ref{fig:height-line}(c). Since a shift of the height $h$
by $3$ corresponds to a string translation by the mean string
distance $\ell$, the displacement on each non-vertical bond is
given by $u= \ell \bar h/3 + u_0$ where $\bar h$ is the coincident
(original) height of the two hexagonal plaquettes which are joined
by the non-vertical bond, see Fig.~\ref{fig:height-line}(a). $u_0$
is a global offset of the flat string lattice.  For $u_0=n \,
a/2$, $n \in {\mathbb Z}$, the class $(++-)$ is chosen while
$u_0=a/4 + n \, a/2$ selects $(+-0)$. This definition of $u$ is
equivalent to the height profile introduced by Zeng and Henley
\cite{Zeng+97} by averaging over the three sites of every triangle
in order to obtain a coarse-grained height on the center of the
triangles.

After a coarse-graining over length scales large compared to the
lattice constant, one obtains a continuous field $u({\bf r})$ which
allows to write the effective {\it free} energy of long-wavelength
fluctuations of the string lattice in the form of a continuum elastic
energy,
\begin{equation}
\label{eq:F-elastic} \frac{{\cal F}_{\rm el}}{T_l}=\int d^2{\bf r}
\left\{\frac {c_{11}}2(\partial_x u)^2 +
\frac{c_{44}}2(\partial_y u)^2 + V_L(u)\right\}
\end{equation}
with compression $c_{11}$ and tilt $c_{44}$ modulus. The tilt
modulus is fixed by the line tension $g$ as $c_{44}=(g/T_l)\rho$.
The compression modulus on asymptotically large length scales is
affected by the entropic repulsion between strings and has to be
determined in a macroscopic way. The compressibility of the
lattice is given by the second derivative of the free energy
density with respect to the mean distance $\ell=1/\rho$ between
the strings \cite{landau1}, which leads to the result
\begin{equation}
\label{eq:c11-from-f}
c_{11}=\ell \frac{\partial^{2}}{\partial \ell^2}[\ell f_l(\ell)]
=\pi^2 \frac{T_l}{g} \rho^3\, ,
\end{equation}
where we have used Eq.(\ref{eq:f_strings}).  $V_L(u)$ is a periodic
potential which reflects the discreteness of the lattice and favors an
ordered phase with flat strings.  Since equivalent flat states are
related by shifts of all straight strings by $a/2$ and since $u=\ell
h/3+u_0$, the locking potential must have the form \cite{footnote1}
\begin{equation}
\label{eq:lock-in-potential} V_L
=-v \cos\left(\frac{4\pi}{a} (u-u_0)\right)
=-v \cos\left(\frac{4\pi}{3} \frac{\ell}{a} h\right)
\end{equation}
with $v>0$.  If $\sqrt{c_{11}c_{44}}>2\pi/a^2$ the string lattice
is sufficiently stiff and the potential $V_L$ is relevant (under
renormalization) so that the strings lock into one of the flat
states. Interestingly, $\sqrt{c_{11}c_{44}}=\pi/\ell^2$ is
independent of the line tension, and the locking potential is
relevant for $\ell < a/\sqrt{2}$. However, the minimal string
separation is $\ell=3a/2$ (corresponding to isotropic dimer
weights), and hence $V_L$ is irrelevant at all possible densities.

However, if there are additional interactions present which
increase the stiffness of the string lattice, the lock-in
potential might become relevant. In fact, we will see below that a
coupling of many TIAF can render the periodic pinning relevant.
Therefore, it is important to determine the consequences of the
lock-in potential for the spin configurations, which depend on the
sign of that potential. To do so, we recall that the offset $u_0$
selects the lattice positions of the strings in the flat states.
It has to be determined from additional information which is in
the present case provided by spin configurations which correspond
to flat strings and small fluctuations around the locked-in
states. Thus, we focus now on the isotropic TIAF so that the
possible classes of flat states are those shown in
Fig.~\ref{fig:flat}.  If one switches between the two classes,
$u_0$ shifts by $a/4$ and thus the sign of $V_L$ changes. The sign
can be determined as follows.

The systems selects the class with the largest entropy, i.e., with the
maximal number of configurations which yield at arbitrary large ${\bf
  r}$ still a finite displacement $\langle [u({\bf r})-u({\bf 0})]^2
\rangle_\text{flat} \sim \text{const.}$, where the displacement is
measured relative to the respective flat state.  For large systems,
the entropy of such macroscopically flat states can be estimated from
the number of string configurations on the hexagonal lattice for which
each string fluctuates in a tube of width $\ell$ about its straight
reference position, i.e., $|u| \le \ell/2=3a/4$ for each non-vertical
dimer \cite{footnote2}. This puts a strong constraint on the spin
states since a flip of a single flippable spin (a spin with 3 up and 3
down spins as neighbors) shifts a string by the hexagonal plaquette on
which the spin sits.  In the class $(+-0)$ the spins on one sublattice
(dark grey in Fig.~\ref{fig:flat}) can assume any of the $2^{N/3}$
possible states where $N$ is the number of triangular lattice sites.
The maximal displacement of $u=3a/4$ is obtained if spins on the other
two sublattices are also flipped with respect to the perfectly ordered
state $(++-)$ which, however, is possible only for $1/2^3 \times 2N/3$
sites since the $3$ neighbors on the fully flippable sublattice must
have the same orientation as the spin to be flipped in order to have
directed and non-crossing strings.  This yields
$2^{N/3+N/12}=\exp(0.2888 \, N)$ configurations. For the fluctuations
in class $(++-)$ all spins on one sublattice (white in
Fig.~\ref{fig:flat}) are frozen.  The sites of the other two
sublattices can be divided into $N/12$ ``rings'', each of $6$
plaquettes centered about a frozen site, and $N/6$ extra plaquettes
between the rings. Due to the constraints on the strings, each ring
permits only $18$ spin states. For a given spin state on all rings,
the extra plaquettes can only flip if its $3$ neighboring spins
located on rings point up, which occurs with probability $(13/18)^3$.
Thus there are $18^{N/12} \times 2^{(13/18)^3 N/6} = \exp(0.2844\, N)$
configurations. We conclude that entropy favors flat states of the
class $(+-0)$ and one has to set $u_0=-a/4$ as shown in
Fig.~\ref{fig:height-line} (c), i.e., $V_L(u)=v\cos(4\pi u/a)$ with
positive amplitude $v$.

\subsection{Spin-spin correlations}
\label{sec:spin-spin-corr}

In this section, we are going to apply the above framework to the spin
correlations of the TIAF model at zero temperature. We concentrate on
isotropic couplings since otherwise the system is ordered. We have
seen that the ground state manifold of the isotropic TIAF can be
regarded as the configuration space of a string lattice. Based on the
irrelevance of lattice pinning effects, one can expect from the
effective elastic description of Eq.(\ref{eq:F-elastic}) an algebraic
decay of correlations which is characteristic for such kind of
two-dimensional systems. In fact, this corresponds to the exact
result found by Stephenson \cite{Stephenson70}.

The displacement correlations of the string lattice can be easily
computed for large ${\bf r}$ from Eq.(\ref{eq:F-elastic}) with the
result
\begin{equation}
\langle(u({\bf r})-u({\bf 0}))^{2}\rangle
=\frac1{\pi}\frac{1}{\sqrt{c_{11}c_{44}}}
\log\left(  \frac{r}{a}\right) \, .
\label{eq:u-correlations}
\end{equation}
So far, we have constructed the string displacement field and the
height profile from a given dimer covering of the hexagonal lattice.
Although the dimer covering in turn is uniquely determined by the
corresponding spin state, we have not yet established a direct
relation between the spin variables $\sigma_i$ and the displacement
field $u$. We will first express the spin variable in terms of the
height profile. For a given spin state, we chose one up spin and
consider its site as the point of origin with height $h=0$. The
construction of the dimer covering from the spin state means that the
height on all other sites is then fixed since parallel (antiparallel)
spins imply a height change by $+2$ ($-1$) from site to site if the
down pointing triangles are traversed clockwise. If a single spin is
flipped, the height on that site changes by $\pm 3$. From this is it
clear that the height profile can locally be changed by $6$ without
affecting the spin state. However, a spin flip with such a change in
$h$ leads out of the ground state manifold. Allowing for this
excitations, the height profile is no longer single-valued. This
induces topological defects which are vortex-anti-vortex pairs in form
of two triangles, each having three parallel spins, see Fig.
\ref{fig:loop}.  Along a closed curve around a single (anti-)vortex
the height changes by $6$.  One easily proves that the above
constraints are met by the spin-height relation
\begin{equation}
\label{eq:h-to-spin}
\sigma_{i}=\cos\left(  \mathbf{q}\cdot\mathbf{r}_{i}+\frac{\pi}{3}
h_{i}\right) = \pm 1 \, ,
\end{equation}
with $\mathbf{q}=(4\pi/(3a),0)$ and the triangular lattice sites ${\bf
  r}_i$. The relation to the string displacement field is then given
by the coarse grained relation $\bar h = 3(u-u_0)/\ell$ with
$\ell=3a/2$ in the isotropic case considered here. Using these
relations, the spin-spin correlation function can be written as
\begin{equation}
\langle\sigma_{i}\sigma_{j}\rangle
=\cos\left[\frac{4\pi}{3a}(x_i-x_j)\right]
e^{- \frac{1}{2} \frac{\pi^2}{\ell^2}
\langle [u({\bf r}_{i})-u({\bf r}_{j})]^2\rangle } \, ,
\label{spin-correlation}
\end{equation}
where we have used the fact that $u({\bf r})$ is Gaussian and that
there are no topological defects at $T=0$. From
Eq.(\ref{eq:u-correlations}) and $\sqrt{c_{11}c_{44}}=\pi/\ell^2$ it
immediately follows that
\begin{equation}
\left\langle \sigma_{i}\sigma_{j}\right\rangle = \cos\left[
\frac{4\pi}{3a}(x_{i}-x_{j})\right]  \frac{1}{|{\bf r}_i-{\bf r}_j|^{\eta}}
\label{eq:spin-correlation-final}
\end{equation}
with $\eta=1/2$ which is in agreement with the exact result
\cite{Stephenson70}.  The anisotropy factor depends here only on
the $x$-coordinates of the spins since the $x$-axis had been
chosen to coincide with one of the triangular lattice directions.

Finally, we consider the height correlations for general anisotropic
dimer weights. These correlations have been studied in the context of
the triangular SOS model some time ago \cite{Bloete+82,Nienhuis+84}.
In those works, the correlations were obtained from four-spin
correlations of the TIAF in the isotropic case, and in terms of the
Pfaffian method for the anisotropic situation. In contrast, here we
present a simple derivation of the correlations based on our effective
elastic description of the string lattice. The mean squared height
difference between two distant positions increases logarithmically
as the string displacement does. The coefficient $K$ measures the
large-scale stiffness of the height profile,
\begin{equation}
\langle(h({\bf r})-h({\bf 0}))^{2}\rangle=\frac1{\pi K}
\log\left(  \frac{r}{a}\right) \, ,
\end{equation}
By making use of the relation between the height profile and the
string displacement field, one easily gets
\begin{equation}
K=\frac{\ell^2}{9}\,\sqrt{c_{11}c_{44}}=\frac{\pi}9 \, ,
\end{equation}
which is independent of the dimer weights $z_1$ and $z_2$. This
remarkable universality was found already in the exact solution of the
dimer model in Refs.~\onlinecite{Bloete+82,Nienhuis+84}, but the
physical reason for that remained unclear. Our interpretation of the
stiffness $K$ as the geometric mean of the two elastic constants of
the string lattice can explain the universality. By changing the dimer
weights, one tunes both the string density and the string tension $g$.
Naively, one would expect that an increasing $g$ would render the
height profile stiffer. However, there is also a reduction of the
entropic repulsion between the strings which accompanies the reduction
of string fluctuations. Interestingly but effects act together as to
generate universality since $c_{44}\sim g$, $c_{11} \sim 1/g$ and
the density dependent relation between $h$ and $u$.

\subsection{Finite temperature}
\label{sec:finite-t}

\placefigure{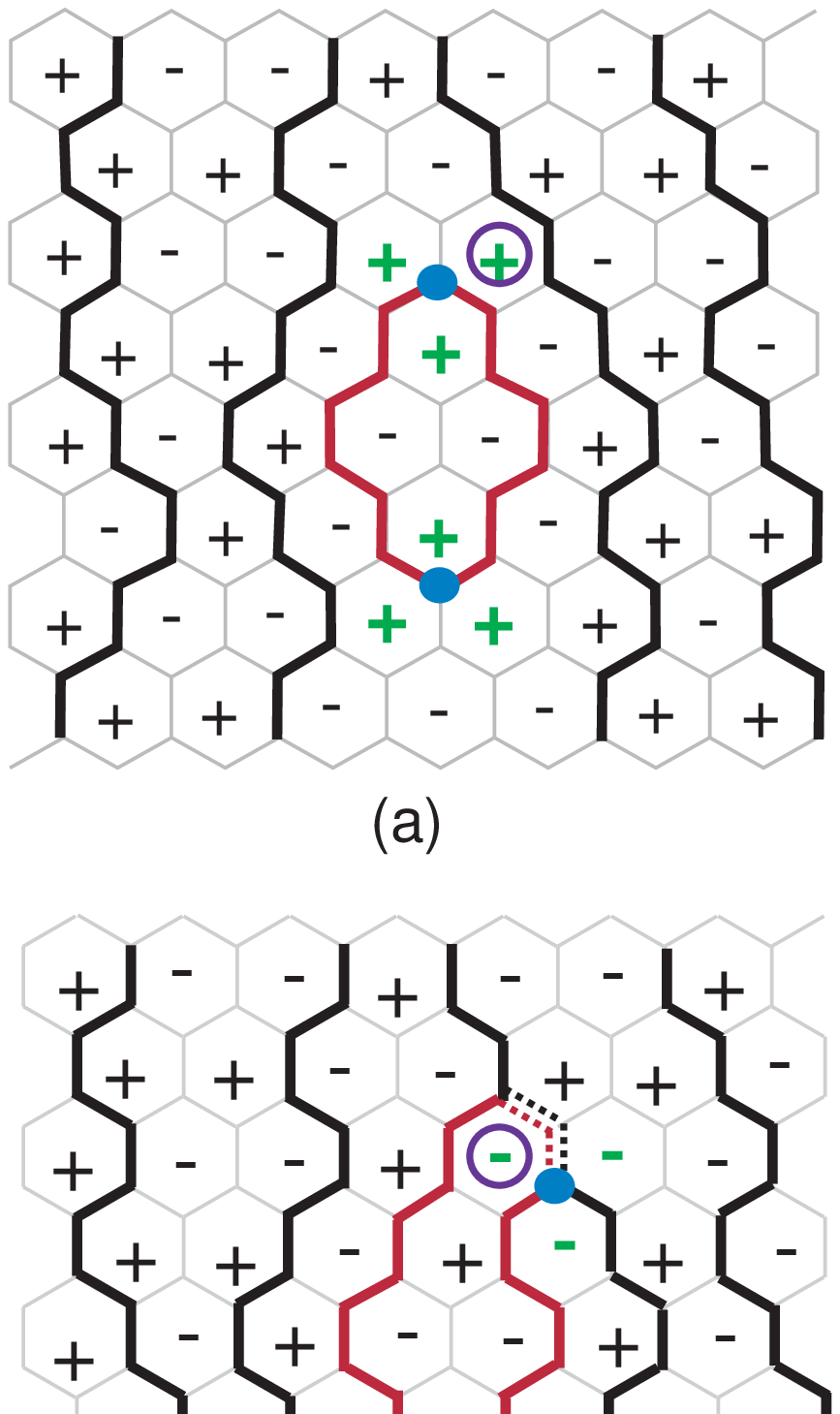}{0.6\linewidth}{(color online) (a)
  Vortex-anti-vortex pair connected by a string loop. (b) Loop
  connected to a string as to generate a non-directed string. This
  configuration is obtained from the one in (a) by flipping a single
  spin (encircled) which costs no energy.}{fig:loop}

For the TIAF at finite temperatures, the mapping from spin states
to string lattices can still be applied. The strings remain
non-crossing, since each triangle can have only one or three
frustrated bonds. However, the triangles with three frustrated
bonds are excitations that generate topological defects, see
Fig.~\ref{fig:loop}(a).  Two defects form a pair that spans a
string loop which is confined between the strings, i.e., strings
cannot cross loops. However, loops can attach to strings resulting
in non-directed strings, see Fig.~\ref{fig:loop}(b).  The fugacity
for the lowest energy single-spin excitations is $e^{-4J/T}$. The
quasi-long-range-order of the $T=0$ phase is destroyed if the
defect pairs can unbind. This will happen at sufficiently weak
string lattice stiffness. Actually, the critical stiffness for
that is $\sqrt{c_{11}c_{44}} = 2\pi/\ell^2$ and hence by a factor
of two larger than the actual stiffness corresponding to the TIAF
at $T=0$.  Hence topological defects are {\it always} a relevant
perturbation but they can occur only for $T>0$ where the fugacity
is finite, leading to exponentially decaying spin correlations.
Thus the system does {\it not} show a Kosterlitz-Thouless
transition by tuning the temperature.

\section{Stacked (2+1)D Ising model}
\label{sec:stacked}

After we have introduced the description of the frustrated 2D TIAF
in terms of fluctuating strings, we shall apply this mapping now
to study the $(2+1)$-dimensional stacked TIAF which consists of
ferromagnetically coupled TIAF layers, see Eq.(\ref{stackedH}). As
in the 2D system, the in-plane frustration is expected to have
strong influence on the underlying physics due to the Ising
symmetry. Actually, the behavior is independent of the interaction
along the stacking direction, whether it is ferro- or
antiferromagnetic. Experimentally, the stacked model is a
reasonable description of triangular cobalt antiferromagnets of
the type ACoX$_3$ where A is an alkali metal and X a halogen atom
\cite{Collins+97}. For these magnets, a strong crystal field
splitting leads to an effective spin $1/2$ state with the moment
oriented along the stacking direction. In these compounds the
in-plane exchange coupling $J_\|$ is small compared to the
inter-plane coupling $J_\perp$. The increased dimensionality of
the stacked system reduces the degeneracy of the ground state so
that the entropy per spin vanishes at $T=0$ and ordering at even
finite-temperature might be possible. This is also suggested by
the following argument \cite{Blankschtein+84}. Each ground state
of the 2D TIAF yields a ground state of the stacked system if all
spins are aligned along the stacking direction, i.e., the
configuration in each layer is identical. Hence, at $T=0$ the spin
correlation function cannot decay in-plane faster than that of the
2D system. However, one might wonder if the system can gain
entropy by introducing domain walls parallel to the layers which
would destroy the order along the stacking direction. The average
energy cost for such a wall is at least
\begin{equation}
\label{eq:dw-argument}
E_{\rm dw} \sim 2J_\perp \int_0^{N^{1/3}} \frac{2\pi r}{\sqrt{r}} dr
\sim N^{1/2}
\end{equation}
where $N$ is the total number of spins of the 3D system and we used
the power law of Eq.~(\ref{eq:spin-correlation-final}). This has to be
compared to the entropy gain which growth only $\sim (1/3) \log N$
since the wall can be placed at any of the $N^{1/3}$ layers. This
naive argument suggests the absence of domain walls at $T=0$, and
order along the stacking direction. Moreover, as we will see more
clearly below, the huge in-plane degeneracy can actually induce order
even in-plane, a phenomenon known as {\it order from disorder}
\cite{Villain+80}.

The stacked TIAF has been studied first by Blankschtein {\it et al}
\cite{Blankschtein+84}. Their results based on the LGW approach and
Monte Carlo simulations suggested the existence of two different
ordered phases and a XY-like transition into the paramagnetic phase.
The LGW Hamiltonian was constructed for the large scale fluctuations
about the two minimal energy modes with wave vectors ${\bf Q}_\pm=(\pm
4\pi/3,0)$, leading to a 3D XY model with a 6-fold symmetry breaking
term,
\begin{eqnarray}
\label{eq:H-LGW}
H_{{ \rm LGW}}(\psi) &=& \int d^3 {\bf r} \Big\{
\frac{1}{2} |\nabla \psi_0|^2
+\frac{r}{2}|\psi_0|^2 + u_4 |\psi_0|^4 \nonumber\\
&&\, + \, u_6|\psi_0|^6 + v_6|\psi_0|^6 \cos\left(6\phi\right)
\Big\},
\end{eqnarray}
for the complex order parameter $\psi({\bf r})=\psi_0 \,
e^{i\phi}$. Since for $p \gtrsim 3.4$ the symmetry breaking term
is irrelevant in 3D [\onlinecite{Aharony+86}], the LGW theory
predicts for the transition to the paramagnetic phase XY
universality \cite{Blankschtein+84,Netz+berker}.  The sign of
$v_6$ is not fixed in this approach, and hence two ordered phases
with a relevant symmetry breaking term are possible in principle
(see below for the two types of ordering). In fact, the two
corresponding phases were observed with increasing temperature in
the Monte Carlo simulations in Ref. \onlinecite{Blankschtein+84}.
However, more recent simulations indicated the existence of only
one ordered phase which corresponds to the one found in Ref.
\onlinecite{Blankschtein+84} at higher temperatures only
\cite{Matsubara+87,Kim+90,Nagai+94,Plumer+95,Moessner+01b}.  This
conclusion is also supported by a hard-spin mean field theory
\cite{Akguc+95}. There is also a controversy about the nature of
the transition to the paramagnetic state. While the simulations of
Heinonen {\it et al.}  indicate tricritical behavior
\cite{Heinonen+89}, a histogram Monte Carlo analysis supports
XY-like behavior \cite{Bunker+93}. These conflicting results have
to be viewed against the background of Coppersmith's work
\cite{Coppersmith85}. She argued that the above mentioned LGW
approach is not reliable since it ignores the geometric
frustration and hence does not yield the correct low temperature
state.

In the following, by making use of the string mapping established
in the preceding section, we provide a microscopic derivation of
the LGW action for the stacked TIAF. This will allow us to discuss
the nature of the ordered phase and the transition to the
paramagnetic state.  In each layer, we relate the spin variables
$\sigma_{ik}$ to a height profile $h_{ik}$ as in the preceding
section so that the relation of Eq.~(\ref{eq:h-to-spin}) holds for
every layer. Then the intra-layer coupling can be written as
\begin{equation}
\label{eq:in-plane-sigma-to-h}
\sigma_{ik}\sigma_{jk}=-\cos\left[\frac{\pi}{3}\left(h_{ik}-h_{jk}+
\eta_{ij}\right)\right]
\end{equation}
with a shift $\eta_{ij}=+1$ for the bond directions $(a,0)$ and
$\eta_{ij}=-1$ for the directions $(a/2,\pm\sqrt{3}a/2)$.  Next, we
have to define how the height should change along the stacking
direction. Since $h_{ik}$ changes by $\pm 3$ for a spin flip, we set
along the column passing through the origin of each layer $h_{0k}=0$
$(3)$ if $\sigma_{0k}=+1$ $(-1)$, where $k$ numbers the layers.
According to this rule, the inter-layer coupling reads
\begin{equation}
\label{eq:out-plane-sigma-to-h}
\sigma_{ik}\sigma_{ik+1}=-\cos\left[\frac{\pi}{3}\left(h_{ik}-h_{ik+1}
\right)\right] \, .
\end{equation}
Now, the height profile can again be interpreted as the
displacement field of strings, which now form a 3-dimensional
lattice. Hence, it is expected to more stable against
fluctuations, and spin order will occur.  However, the periodic
couplings of the displacement field allow for topological defects
which eventually drive the system into the paramagnetic state.
Even in the presence of defects, the picture of non-crossing
strings remains valid since a triangle can have either 1 or 3
frustrated bonds. When introducing continuous fields, the lock-in
potential of Eq.(\ref{eq:lock-in-potential}) has to be included as
to reflect the original discreteness of the height which reads
$h_{ik}=3(u_{ik}-u_0)/\ell$ in terms of the string displacement
$u_{ik}$. Hence, we obtain the reduced 3D string Hamiltonian
\begin{eqnarray}
\label{eq:classical-strings} \!\!\!\!\!\!\!\!\!\!
H_S\!\!&=&\!\!-\tilde K_\| \!\!\sum_{\langle ij\rangle,k}\!
\cos\left[\frac{\pi}{\ell}(u_{ik}-u_{jk}+\eta_{ij}a/2)\right]
\nonumber \\
&-& \!\! \tilde K_\perp \!\sum_{i,k}\!\cos\left[\frac{\pi}{\ell}
(u_{ik}\!-\!u_{ik+1})\right]
+v\!\sum_{i,k}\!\cos\left[\frac{6\pi}{\ell} u_{ik}\right]\nonumber\\
\end{eqnarray}
with couplings $\tilde K_\|=J_\|/T$, $\tilde K_\perp=J_\perp/T$ and
mean string separation $\ell=3a/2$.  Here $v>0$ due to the entropy
argument at the end of Section~\ref{sec:dimers-to-strings}.  The
in-plane shift $\eta_{ij}$ reflects frustration since $\sum \eta_{ij}
= 3$, where the sum runs over the bonds of a down pointing triangle.
Since in the discrete version the $u_{ik}$ can vary over the bonds
only by $+a$ or $-a/2$, the energy is minimized for a non-uniform
change of $u_{ik}$ along the triangles.  This is distinct from a
priori continuous field of an antiferromagnetic XY model on a
triangular lattice whose oriented uniform change defines a helicity,
giving rise to an additional ${\mathbb Z}_2$ symmetry \cite{leedh1}.
Hence, the latter $XY$ model exhibits a Ising transition because of
the ${\mathbb Z}_2$ symmetry breaking of the ground states
\cite{leedh1}.  However, in the present case, the non-crossing
restriction for the strings, originating from the geometrical
frustration, prohibits this extra symmetry breaking.

Thus the stacked Ising model of Eq.(\ref{stackedH}) maps to a stack of
planar lattices of {\it non-crossing} strings which is described by a
(2+1)D frustrated XY model with a 6-fold clock term.  Interestingly,
this provides a microscopic derivation of the LGW theory
\cite{Blankschtein+84} if the string displacement $u$ is identified
with the phase $\phi$ of the order parameter via $\phi=\pi u/\ell$.
However, our effective model differs from the LGW theory in two
important points: (i) the in-plane XY coupling is frustrated and (ii)
there is a topological constraint on $\phi$ since $u$ is restricted by
the non-crossing condition. The latter point will lead to an
entropicly increased phase stiffness on large length scales.

\placefigure{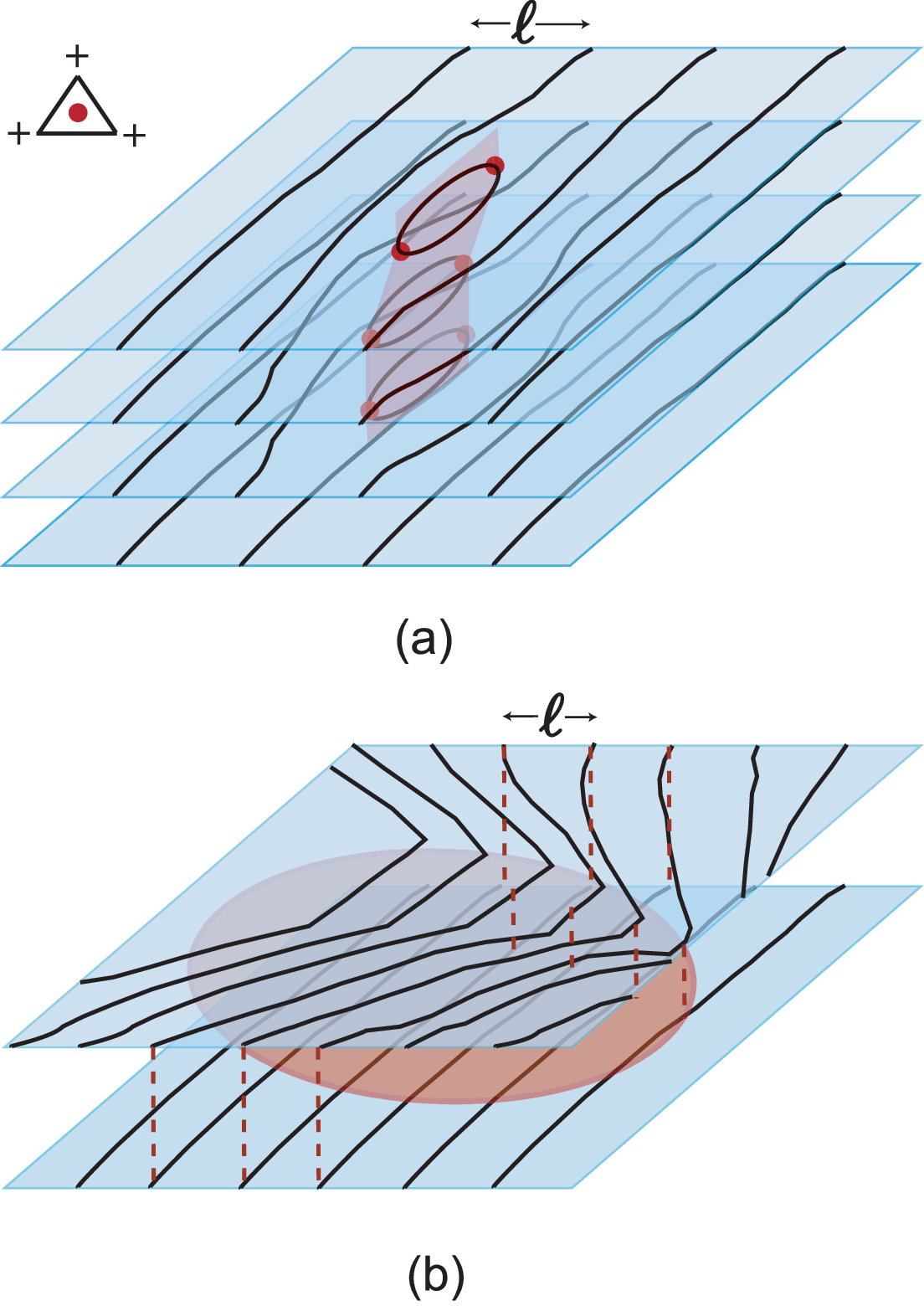}{0.7\linewidth}{(color online) Different type
of topological defects. (a) vortex-anti-vortex pairs in the layers
form vortex loops perpendicular to the planes. (b) vortex loop
parallel to the planes which borders the mismatch regions in which
the strings in two adjacent planes are shifted by $2\ell$.
}{fig:defects}

With the model of Eq.~\eqref{eq:classical-strings} at hand, we can
analyze the ordering mechanism and the transition to the
paramagnetic phase. The latter transition is driven by topological
defects which are generated by thermal fluctuations. It is well
known that the XY coupling allows for defect loops across which
$\phi$ changes by $2\pi$.  These defects are superpositions of two
kind of loops [cf.~Fig.~\ref{fig:defects}]: fully frustrated
triangles with all spins aligned form in-plane vortex-anti-vortex
pairs, which can be viewed as a string forming an in-plane loop.
Together with other pairs in the neighboring layers they form
defect loops oriented perpendicular to the planes,
cf.~Fig.~\ref{fig:defects}(a).  Another type of defect loops is
oriented parallel to the layers.  They arise as the boundaries of
2D areas across which strings in adjacent planes are shifted by
$2\ell$, cf.~Fig.~\ref{fig:defects}(b). For the universality class
of the transition to the ordered state it is important to take
into account the stacked nature of the $(2+1)$D XY model and the
$6$-fold clock term. In turns out that the stacking is irrelevant
since the layers cannot decouple independently from the vortex
unbinding transition within the layers. This is due to the fact
that unbinding of defects oriented parallel to the layers can
occur only at a critical value for the effective coupling $K_\|$
which is by factor of $1/8$ below that for the in-plane
dissociation of vortices \cite{Korshunov90}.  In addition, a
$6$-fold clock term is known to be irrelevant at the XY critical
point under renormalization if $p\gtrsim 3.4$
[\onlinecite{Aharony+86}]. Hence we conclude that the transition
from the paramagnetic phase to an ordered state must be in the 3D
XY universality class.

In the following, let us discuss the property of the ordered
phase. In the latter phase, the XY couplings of
Eq.~(\ref{eq:classical-strings}) can be expanded in $u_{ik}$, and
in the continuum limit each layer is described by
Eq.~(\ref{eq:F-elastic}) with an additional harmonic inter-layer
coupling. Moreover, the $6$-fold clock is relevant, and the
strings are locked into a flat state with spin order whose nature
depends on the form of the locking potential. In order to compare
our approach with the LGW theory \cite{Blankschtein+84}, we
discuss potential ordered states.  As we have seen in Section
\ref{sec:dimers-to-strings}, there are two classes of flat
strings. They differ by their value of the shift $u_0$ in the
lock-in potential $V_L(u)$.  After going over from the discrete
field $u_{ik}$ to a coarse-grained description with a continuous
field $u({\bf r})$, the coarse-grained spin variable can be
written as
\begin{equation}
\label{eq:av-u-to-spin}
\sigma({\bf R}_{jk})=\cos\left(  \mathbf{Q}\cdot\mathbf{R}_{jk}+
\frac{2\pi}{3a} (u({\bf R}_{jk})-u_0)\right) \, ,
\end{equation}
where ${\bf R}_{jk}$ are the 3D lattice sites and ${\bf Q}=
(4\pi/(3a),0,0)$. For flat strings, the coarse-grained field $u({\bf
  r})$ is zero on average, and the global shift $u_0$ determines the
sublattice magnetizations.  For the first state, cf.~left part of
Fig.~\ref{fig:flat}, $u_0=-a/4$, and there is an additional shift of
$\pi/6$ in Eq.(\ref{eq:av-u-to-spin}), leading to the sublattice
magnetizations $\langle \sigma \rangle =
(\frac{\sqrt{3}}{2},-\frac{\sqrt{3}}{2},0)$.  For the second state,
cf.~right part of Fig.~\ref{fig:flat}, $u_0=0$, one has the sublattice
magnetizations $\langle \sigma \rangle =
(1,-\frac{1}{2},-\frac{1}{2})$ since ${\bf QR}_{jk}=0$, $4\pi/3$, or
$-4\pi/3$ on the three sublattices. Hence all sublattices are at least
partially ordered.  Since our entropy argument lead to $u_0=-a/4$, the
first state with one fully disordered sublattice should set up the
ordered phase at all temperatures which appears to be consistent with
more recent simulations.

How can this ordering be reconciled with the fact that each ground
state of the highly degenerate and hence disordered 2D TIAF is also a
ground state of the stacked model when all spins are aligned along the
stacking direction? This is the point where the mechanism of {\it
  order from disorder} comes into play. As we have seen at the
beginning of this section, the formation of domain walls parallel to
the layers is energetically not favorable. However, the system can
gain entropy by flipping {\it single} (so called flippable) spins in a
single layer which does not cost energy within the layer. Due to the
huge ground state degeneracy there are many configurations which just
differ in their orientation of flippable spins. Hence, a subset of all
the 2D ground states is selected which allows for those single spin
flips. This explains the existence of the found ordered state since it
is composed only of an appropriate subset of all 2D ground states
which on average has one sublattice disordered. The spins on this
disordered sublattice form chains along the stacking directions which
are decoupled from each other at low temperatures, and hence should
behave as individual 1D Ising spin chains.  At sufficiently low
temperatures, the stacked system is thus expected to show excitations
which are 1D Ising-like.  This conclusion is consistent with
multi-spin Monte Carlo simulations of the specific heat at low
temperatures \cite{Kim+90}.

\section{Quantum frustrated model}
\label{sec:quantum}

Having established the relation of the $(2+1)$-dimensional TIAF to a
lattice of fluctuating strings with topological defects, we now will
apply this relation to the $2$-dimensional TIAF in a transverse
magnetic field. The Hamiltonian of this model is given in
Eq.~\eqref{eq:model}. The transverse field introduces simple quantum
dynamics to the highly degenerate critical state of the classical
TIAF. Due to the general relation between 2D quantum spin systems and
3D classical Ising spin models \cite{suzuki1}, one can expect from the
results for the stacked TIAF quantum order arising from the interplay
of quantum fluctuations and geometric frustration. The existence of a
quantum ordered state was suggested in recent works on the basis of a
LGW theory \cite{moessner1}, a kink model \cite{Mostovoy+03}, and
simulations \cite{Isakov+03}.

The exact mapping from the 2D quantum TIAF to a classical stacked
Ising system is provided by Suzuki's theorem which states that the
partition function of the quantum system can be written as
\cite{suzuki1}
\begin{eqnarray}
Z  &=&\mathrm{Tr}\;\exp\left\{  -\frac{J}{T}\sum_{\langle
i,j\rangle}\sigma_{i} ^{z}
\sigma_{j} ^{z} +\frac{\Gamma}{T}\sum_{i} \sigma_{i} ^{x}\right\} \nonumber\\
&=&\lim_{n\to\infty} \left[  \!\frac12
\sinh\left(\frac{2
\Gamma}{nT}\right)\right]^{\frac{nN}{2}}
\!\!\sum_{\{\sigma_{ik}\}}
\!\exp\Big\{\!-\frac{J}{nT}\!\!\sum_{\langle i,j\rangle,k}
\!\sigma_{ik}\sigma_{jk} \nonumber\\
&+& \frac{1}{2} \log\coth\left(\frac
{\Gamma}{nT}\right)\sum_{i,k}\sigma_{i,k}\sigma_{i,k+1}\Big\}
, \label{eq:suzuki}
\end{eqnarray}
where $\sigma_{ik}=\pm 1$ are classical Ising variables which are
defined on a stacked system of $n$ antiferromagnetic triangular
lattices, and the second index $k=1,\ldots,n$ refers to the layer
number. Hence, in the relevant limit of large $n$ the quantum system
is described by the Hamiltonian of the stacked TIAF of
Eq.~(\ref{stackedH}) with the couplings $J_\|$ and $J_\perp$ replaced
by a infinitesimal small in-plane coupling $\tilde K_\|=J/(nT)$ and a
strong ferromagnetic inter-layer coupling $\tilde
K_\perp=\frac{1}{2}\log(nT/\Gamma)$ which is controlled by the strength
of the quantum fluctuations, i.e., the transverse field $\Gamma$.  In
analogy with the preceding section, the quantum system is hence
described by a $(2+1)$D string lattice with the Hamiltonian of
Eq.~(\ref{eq:classical-strings}).  This suggests that, as for the
classical 3D system, the transverse field Ising model must have an
ordered phase at small transverse field and a quantum phase transition
in the 3D XY universality class to a disordered state. This scenario
was previously predicted from by the LGW approach. However, in
contrast to the latter approach, our mapping of the quantum frustrated
spin system to strings yields even the coupling constants of
Eq.~(\ref{eq:classical-strings}) which is of utmost importance for the
construction of the phase diagram.

\subsection{Quantum critical point}

First, let us focus on the quantum phase transition to the
paramagnetic phase. The purpose of the following analysis is to
estimate the location of the critical point and to show that the
Trotter limit $n\to\infty$ can be carried out explicitly in our
approach. It is useful to separate the partition function associated
with the XY model of Eq.~(\ref{eq:classical-strings}) into a spin wave
part and a vortex part. This can be done by making use of the fact
that the planar model can be mapped to a simpler model proposed by
Villain \cite{Villain75}. As was shown by Kleinert \cite{kleinert1}
and in Ref.~\onlinecite{jose1}, in the partition function one
can make the substitution
\begin{equation}
e^{\tilde K \cos(\phi)} \to c \! \sum_{m=-\infty}^\infty
e^{-K(\phi-2\pi m)^2/2}\, ,
\end{equation}
where the right hand side is known as Villain coupling
\cite{kleinert1,jose1} with a new $\tilde K$-dependent coupling
$K$ .  Both expressions become identical in the two limits of
$\tilde K \to 0$ and $\tilde K \to \infty$ which are fortunately
precisely the two cases arising from the Trotter limit
$n\to\infty$. Then the coupling constant $K$ and the coefficient
$c$ are given by
\begin{eqnarray}
K = \tilde K, & \, & c = e^{K} \quad
\text{for} \quad \tilde K\to \infty \\
K = \frac{1}{2\log (2/\tilde K)}, & \, & c=\sqrt{2\pi K} \quad
\text{for} \quad \tilde K\to 0
\end{eqnarray}
Hence, after taking the Trotter limit, the coupling constants of
the equivalent Villain model both scale logarithmically
with $nT$,
\begin{equation}
\label{eq:Villain-Ks} K_\perp = \frac{1}{2}
\log\left(\frac{nT}{\Gamma}\right), \quad
K_\| = \frac{1}{2}\frac{1}{\log(2nT/J)}\, .
\end{equation}
Only for $T=0$, one can set $T\sim J/n$ so that for $n \to \infty$ the
coupling $K_\perp$ remains finite, and the system shows 3D
behavior. For any finite $T$, however, $K_\perp$ must diverge with $n$
and the system shows a 3D to 2D crossover with increasing length
scales.

Knowing the coupling constants of the Villain model with decoupled
spin wave and vortex parts for large $n$, we can go ahead and apply
dimensional crossover scaling in order to make quantitative
predictions about the location and universality of the quantum
critical point. We start by eliminating $n$ from
Eq.~(\ref{eq:Villain-Ks}).  From this we obtain the temperature
independent relation
\begin{equation}
\label{eq:exact-rel-Ks}
  K_\perp = \frac{1}{4K_\|} - \frac{1}{2}\,
\log \left( 2\,\frac{\Gamma}{J}\right)
\end{equation}
between $K_\|$ and $K_\perp$ which makes the parameter space one
dimensional. The geometric mean of the coupling constants controls the
stiffness of the XY system. It can be expanded in the relevant limit
of small $K_\|$ as
\begin{equation}
\label{eq:rel-between-Ks} \sqrt{K_\| K_\perp} =\frac{1}{2}
\left[1-\log(2\Gamma/J) \, K_\| + {\cal O}(K_\|^2)\right] \, ,
\end{equation}
where we used the relation of Eq.~\eqref{eq:exact-rel-Ks}.  The latter
relation will be compared to a crossover scaling result which provides
a quantitative description of the increase of the transition
temperature of a layered (2+1)D XY model under the increase of the
number of layers. Following closely the analysis in
Refs. \onlinecite{Ambegaokar+80,schneider1}, one obtains a relation
between the 3D critical value $K_\infty^c$ for the in-plane coupling
and the corresponding $K_n^c$ for a system of $n$ layers,
\begin{equation}
\label{eq:dim-cross}
\frac{1}{n}\frac{K_\infty^{c}}{K_n^{c}}=\gamma
\left(\frac{K_\|}{K_{\perp}}\right)^{1/2}
\left(1-\frac{K_\infty^{c}}{K_n^{c}}\right)^\nu
\end{equation}
with the critical exponent $\nu\approx 2/3$ of the 3D XY model and a
numerical constant $\gamma$. For $n=1$ this relation yields at
the 3D critical point with $K_\|=K_\infty^c$ the following expression
for the geometric mean of the coupling constants,
\begin{eqnarray}
\label{eq:3d-k}
\sqrt{K_\infty^c K_\perp} &=&
\gamma K_1^c
\left( 1 - \frac{K_\infty^c}{K_1^c} \right)^\nu \nonumber\\
&=& \gamma K_1^c
\left( 1 - \nu \frac{K_\infty^c}{K_1^c} + \dots\right)
\, ,
\end{eqnarray}
where we have expanded for $K_\infty^c \ll K_1^c$. For consistency,
the latter expression should be of the same form as the expression in
Eq.~\eqref{eq:rel-between-Ks} at the quantum critical point with
$\Gamma=\Gamma_c$ corresponding to $K_\|=K_\infty^c$. Indeed, both
expressions are identical if one sets
\begin{equation}
\gamma=1/(2K_1^c),\quad 2\Gamma_c/J=e^{\nu/K_1^c}.
\end{equation}
Hence, the location of the quantum critical point is determined by the
critical value for the coupling of the 2D XY model on a triangular
lattice. The standard Kosterlitz-Thouless argument for the vortex
unbinding transition yields \cite{leedh1} in that case $K_1^c=2/\pi
\times 2/\sqrt{3}$, leading to $\Gamma_c/J=1.24$.  However, this
scaling approach neglects renormalization effects due to the $6$-fold
clock term and the non-crossing of strings which should provide a net
increase of $\Gamma_c$. This is consistent with recent Monte Carlo
studies which suggest $\Gamma_c/J\approx 1.65$
[\onlinecite{Isakov+03}].

\subsection{Phase diagram}

\placefigure{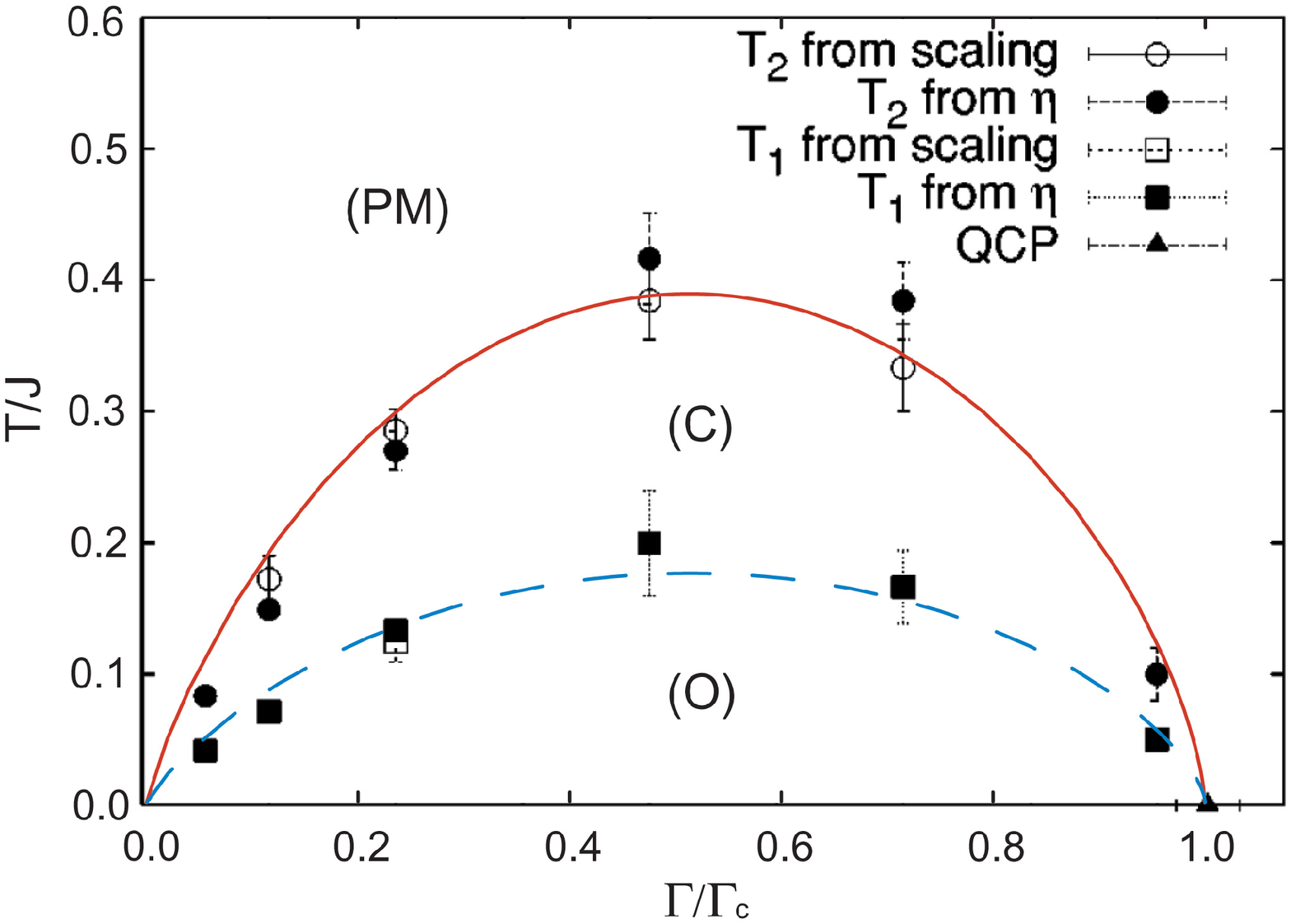}{0.9\linewidth}{(color online) Phase diagram
  with the phase boundaries given by Eq.~(\ref{eq:boundary}) with
  $b=0.98$ and Monte Carlo results of Fig.~1 in Ref.
  \onlinecite{Isakov+03}. A critical phase (C) is separated by a
  boundary at $T_{c,2}$ from the paramagnetic phase (PM) and at
  $T_{c,1}=4/9\, T_{c,2}$ from an ordered phase (O).
}{fig:phasediagram}

Having established the existence of an ordered phase for
$\Gamma<\Gamma_c$ at $T=0$, it is important to study the stability of
this phase against thermal fluctuations. For finite temperatures, 2D
XY physics should dominate at large length scales, and two
Kosterlitz-Thouless transitions separating a critical phase at
intermediate temperatures from the ordered and the paramagnetic state,
respectively, are expected \cite{jose1}. In the following we will
construct a quantitative phase diagram for the transverse field TIAF
in the $\Gamma$-$T$-plane.  The stability range of the phase with
bound defects can be estimated from the crossover scaling formula of
Eq.~(\ref{eq:dim-cross}). Setting in the latter expression $K_\|$ to
its critical value $K_n^c$ for a $n$-layer system, and using the
scaling formula also at $n=1$ in order to relate $K_\infty^c$ to
$K_1^c$ and the bare coupling ratio $K_\perp/K_\|$, we can derive an
expression for the Kosterlitz-Thouless transition temperature
$T_{c,2}(\Gamma)$ of the critical phase (C) which has only bound
defects [see Appendix \ref{sec:appedix} for details],
\begin{equation}
\label{eq:boundary} \frac{T_{c,2}}{J} = b \,
\frac{\Gamma}{\Gamma_c} \, \log^\nu \left( \frac{\Gamma_c}{\Gamma}
\right)  \, ,
\end{equation}
where $b$ is a numerical constant which is fixed by the (unknown)
renormalization of $K_\perp$ and remains finite for $n \to
\infty$. At large but finite $n$, for consistency, the
renormalized effective coupling for the $n$-layer system must
behave at as $K_\perp^{\rm eff} \sim (n/n_c)^{3/5}$ with a
characteristic number $n_c=bK_1^c (J/T) (\Gamma/\Gamma_c)$ which
is a measure for the strength of quantum fluctuations and
characterizes the effective system size along the Trotter
(``imaginary time'') axis.  The spin correlation function in the
critical phase (C) decays according to
Eq.~(\ref{eq:spin-correlation-final}) with the exponent $\eta$
varying continuously between $\eta=1/4$ at $T_{c,2}$ and
$\eta=1/9$ at a lower critical temperature $T_{c,1}$ which marks
the transition to the ordered phase (O) \cite{jose1}. The
correlation function exponent behaves discontinuously at
$\Gamma=0$ where it is $\eta=1/2$ and at the quantum critical
point where the 3D XY result $\eta\approx 0.040$ holds
\cite{Zinn-Justin}.

At the lower critical temperature $T_{c,1}$ there is a second
Kosterlitz-Thouless transition to an ordered state where the
clock-term in Eq.~(\ref{eq:classical-strings}) is relevant and
locks the strings to the lattice. For a single layer, this
transition occurs at the critical coupling $K_1^c=9/(2\pi) \times
2/\sqrt{3}$, see Ref.~\onlinecite{jose1}. The boundary of the
ordered phase can be obtained analogously to that at $T_{c,2}$,
yielding
\begin{equation}
T_{c,1}=\frac{4}{9} \, T_{c,2} \, .
\end{equation}
Close to the quantum critical point both Kosterlitz-Thouless transition
temperatures vanish $\sim (\Gamma_c - \Gamma)^\nu$ as expected from
scaling. Fig.~\ref{fig:phasediagram} compares our analytical results
to recent Monte Carlo data for the phase boundaries, showing very good
agreement across the entire range of $\Gamma$ if we set $b=0.98$ in
Eq.(\ref{eq:boundary}). According to our analysis of the ordered phase
for the stacked system in Section \ref{sec:stacked} the quantum
ordered phase (O) is characterized by the finite sublattice
magnetizations $(\sqrt{3}/2,-\sqrt{3}/2,0)$.  This type of order is
consistent with recent simulations \cite{Isakov+03}.

\section{Conclusion and Extensions}
\label{sec:discussion}

In this article, we have studied for a 2D Ising system the
combined effect of classic geometric frustration and an ordering
mechanism which either results from stacking many 2D system or by
allowing for quantum dynamics from a transverse field. We have
presented an exact relation between spin variables and the
displacement field of a string lattice which we used to derive an
effective Hamiltonian which was obtained previously only on the
basis of a LGW approach. This allowed us to determine the nature
of the ordered phase, the universality of the quantum phase
transition to the paramagnetic phase, and the quantitative phase
diagram at finite temperatures. To our knowledge, this is the
first approach for the analyzed models which explicitly performs
the Trotter limit $n\to \infty$ to obtain quantitative results for
the transition temperatures. We have compared the so obtained
phase diagram to predictions of recent simulations and found
satisfying agreement. We related the spin model also to its dimer
model representation and the resulting height profile description.
Implementing a simple random walk argument for the strings, we
could derive the exact free energy in the continuum limit which
was known before only from a more complicated Pfaffian method. Our
approach could also explain the physical reason for the previously
found universality of the large-scale stiffness of the height
profile in the presence of anisotropic bond weights.

For future extensions of our approach it is interesting to note that
the string analogy applies independently of the form of the spin
couplings. Interestingly, the mapping to strings should also be useful
in the presence of quenched disorder when the system usually exhibits
spin glass behavior. The behavior of a 2D string lattice with pinning
sites \cite{emig1} suggests that in geometrically frustrated systems
the spin glass correlation function \cite{Binder-Young-RMP}
$g_{SG}\left( {\bf r}_{i}-{\bf r}_{j}\right) =\overline{<\sigma_{i}
  \sigma_{j}>^{2}}$ decays at least according to a power law, which
implies that the spin glass phase is destroyed by the huge degeneracy
from the geometric frustration. However, quenched disorder will induce
topological defects, and a full analysis is more complicated. The
situation is under better control for geometrically frustrated Ising
systems with random dilution. The dilution sites act as pinning
centers for the strings and topological defects do not arise from this
type of disorder.

In the quantum frustrated system, though it is very hard to get a
clear picture of how the transverse field and quenched disorder act
together, one can try to make use of a method analogous with what
Moessner \textit{et al.} \cite{moessner1} used in discussing the
existence of an ordered phase in clean quantum frustrated Ising
systems. Moessner \textit{et al.} used the divergence of the
integrated squared two-spin correlation function to argue about a
non-analytic free energy. Instead, with quenched disorder, the spin
glass correlation function had to be integrated to detect a
singularity in the disorder averaged free energy. The divergence of
the latter integral would then signal the possibility of an (spin
glass) ordered phase, induced by quantum fluctuations.

Finally, we discuss potential implications for the quantum phase
transition in 2D Ising spin glass models in a transverse field
\cite{Rieger-Young-PRL}.  From the analysis in Section
\ref{sec:quantum}, we know that the 2D quantum frustrated Ising
model can be mapped to a system of classical stacked string
lattices which is described by a 3D $XY$ Hamiltonian.  With
quenched randomness, the quantum spin system should be mapped to a
3D $XY$ model with \textit{columnar}, i.e., correlated disorder.
If we assume that the lattice type (and hence the geometric
frustration) is irrelevant for the universality class of the
transition, we can argue that the quantum phase transition of 2D
transverse field Ising spin glasses should belong to the
universality class of 3D $XY$ model with columnar disorder.
Interestingly, this speculation seems to be consistent with recent
numerical results for both a transverse field Ising spin glass
\cite{Rieger-Young-PRL} and a 3D $XY$ model with columnar disorder
\cite{Vestergren-Teitel-Wallin-PRB}. The two corresponding
exponents, the dynamical exponent $z=1.5\pm0.05$ for the spin
glass and the anisotropy exponent $\zeta=1.3\pm0.1$ for the XY
model, are indeed in some agreement.

\section*{Acknowledgment}

We thank H. Rieger, S. Scheidl and S. Bogner for helpful
discussion. This work was supported by the Emmy Noether grant No.
EM70/2-3 and the SFB 608 from the DFG .

\appendix

\section{Phase boundary}
\label{sec:appedix}

In this appendix we derive the boundary of the critical phase (C).  We
start from the scaling result of Eq.~\eqref{eq:dim-cross} with
$n=1$. This allows us to compute a relation between the critical
couplings $K_1^c$ and $K_\infty^c$. In the limit $K_\infty^c \ll
K_1^c$ we find
\begin{equation}
\label{eq:rel-K1-Ki}
K_\infty^c = \frac{K_1^c }{\nu+\gamma^{-1}\sqrt{K_\perp/K_\|}} \, ,
\end{equation}
where $\nu$ is the critical exponent of the 3D XY model, and
$\gamma=1/(2K_1^c)$. Using the latter result again in the scaling
formula of Eq.~\eqref{eq:dim-cross} for arbitrary $n$ and with
$K_\|=K_n^c$ set to its critical value, we obtain
\begin{eqnarray}
\label{eq:K_n}
\frac{1}{K_n^c}
&=&
\frac{n}{K_1^c} \left( \nu\gamma\sqrt
{\frac{K_n^c}{K_\perp}}+1\right)
\nonumber \\
&& \times \left( 1-\frac{K_{1}^{c}}{\nu K_n^c
+\gamma^{-1}\sqrt{K_\perp K_n^c}}\right)^{\nu}\, ,
\end{eqnarray}
which yields $K_n^c$ as a function of the inter-layer coupling
$K_\perp$ at large but finite $n$.  We continue by introducing the
bare coupling constant $\zeta_0= e^{2K_\perp}$ so that $n=\zeta_0
\Gamma/T$.  The renormalized effective coupling constant will be
denoted by $\zeta=e^{2K_\perp^{\rm eff}}$. This notation allows us to
take into account the fact that the renormalization of the couplings
$K_\|$ and $K_\perp$ is dependent due to their relation by
Eq.~\eqref{eq:exact-rel-Ks}. Then the effective couplings can be
written as
\begin{equation}
\label{eq:eff-couplings}
K_\|^{\rm eff}=\frac{1}{2\log(2\zeta\Gamma/J)}, \quad
K_\perp^{\rm eff} = \frac{1}{2} \log \zeta \, .
\end{equation}
Now we use $K_n^c=K_\|^{\rm eff}$ and $K_\perp=K_\perp^{\rm eff}$ in
Eq.~\eqref{eq:K_n}, and expand the resulting equation for large
$\zeta$ which gives
\begin{equation}
1=\frac{\zeta_0}{(2\log \zeta)^{5/3}K_1^c}\frac{\Gamma}{T}
\left(\frac{\nu}{K_1^c}-\log\left(2\frac{\Gamma}{J}\right) \right)^\nu\, .
\end{equation}
Multiplying with $T/J$ and using the relation $2\Gamma_c=J
e^{\nu/K_1^c}$ for the quantum critical point, we get the final result
\begin{equation}
  \frac{T_{c,2}}{J} =
  \frac{\zeta_0 e^{\nu/K_1^c}}{2K_1^c (2\log \zeta)^{5/3}}
  \frac{\Gamma}{\Gamma_c} \, \log^\nu \left( \frac{\Gamma_c}{\Gamma}
  \right)  \, ,
\end{equation}
which corresponds to Eq.~\eqref{eq:boundary} if we denote by $b$ the
coefficient in the latter expression. For consistency, it follows from
$K_\perp^{\rm eff} = \frac{1}{2}\log \zeta$ and the definition of $b$
that the effective inter-layer coupling must diverge with $n\to\infty$
according to $K_\perp^{\rm eff}= (n/n_c)^{3/5}$ with a characteristic
$n_c=bK_1^c J\Gamma/(T\Gamma_c)$ which itself diverges for $T\to 0$.


\begin{thebibliography}{99}
\bibitem{footnote1} Since the height $h$ is defined modulo an
  arbitrary (non-integer) global constant $h_0$, the minima of
  $V_L(h)$ can be shifted by $h_0$.  However, $V_L(u)$ remains
  unchanged since the string positions are not effected (modulo $a/2$)
  by such a shift but then $u = \ell (h-h_0)/3 +u_0$.
\end{thebibliography}
\end{document}